\documentclass[12pt]{article}
\usepackage{pdproc,epsfig} 

\begin{document}

\renewcommand{\thefootnote}{\alph{footnote}}

%% to make online citations, e.g."see  Ref. X", instead of superscript
\makeatletter  
 \def\citenum#1{{\def\@cite##1##2{##1}\cite{#1}}}
\makeatother

\title{
NEUTRINOS FROM COSMIC RAY ACCELERATORS\\ IN THE CYGNUS REGION OF THE GALAXY}

\author{FRANCIS HALZEN and AONGUS \'O\,MURCHADHA}

\address{Department of Physics, University of Wisconsin,
Madison, WI 53706, USA\\
 {\rm E-mail: francis.halzen@icecube.wisc.edu}}

\vspace*{-.3cm}

\abstract{While supernova remnants have been identified as the most likely sources of the galactic cosmic rays, no conclusive observational evidence for this association exists. We show here that IceCube has the possibility of producing incontrovertible evidence by detecting neutrinos produced by the cosmic ray beam interacting with the hydrogen in the vicinity of the supernova shock expanding into the interstellar medium. We show that the observational information on gamma ray fluxes from the Cygnus region, although limited, is sufficient to pinpoint the expected event rate of the neutrinos associated with a single source of 0.5 Crab at the TeV level to within a factor of two, between 2 and 3.8 neutrinos per year. Finally, we note that recent gamma-ray observations reveal the presence of at least three and possibly up to eight such sources, raising the possibility of seeing more than 10 neutrinos per year from these sources alone. }
   
\normalsize\baselineskip=15pt

\section{Cosmic Neutrinos Associated with Galactic Cosmic Rays}

Cosmic accelerators produce particles with energies in excess of $10^8$\,TeV; we do not know where or how. The flux of cosmic rays observed at Earth follows a broken power law. The two power laws are separated by a feature dubbed the ``knee"\llap. Circumstantial evidence exists that cosmic rays, up to perhaps EeV energy, originate in galactic supernova remnants. Any association with our Galaxy disappears in the vicinity of a second feature in the spectrum referred to as the ``ankle"\llap. Above the ankle, the gyroradius of a proton in the galactic magnetic field exceeds the size of the Galaxy and it is generally assumed that we are  witnessing the onset of an extragalactic component in the spectrum that extends to energies beyond 100\,EeV. While the enigmatic nature of the highest energy cosmic rays has been widely advertised, it is also a fact that the origin of the galactic cosmic rays has not been established. We will show here that the positive identification of  supernova remnants as the cosmic ray accelerators is possible with the kilometer-scale neutrino observatories now in the planning or construction phase.

Simple energetics point at the supernova origin of the galactic cosmic rays. By integrating the observed flux we can obtain the energy density $\rho_E$ of cosmic rays in the galaxy from the relation that flux${}={}$velocity${}\times{}$density, or
\begin{equation}
4\pi \int  dE \left\{ E{dN\over dE} \right\} =  c\rho_E\,.
\end{equation}
The answer is that $\rho_E \sim 10^{-12}$\,erg\,cm$^{-3}$. This is also the value of the corresponding energy density $B^2/8\pi$ of the microgauss magnetic field in the galaxy. The accelerator power needed to maintain this energy density is $10^{-26}$\,erg/cm$^3$s given that the average containment time of the cosmic rays in our galaxy is $3\times10^6$\,years. For a nominal volume of the galactic disk of $10^{67}$\,cm$^3$ this requires an accelerator delivering $10^{41}$\,erg/s. This happens to be 10\% of the power produced by supernovae releasing $10 ^{51}\,$erg every 30 years. This coincidence is the basis for the idea that shocks produced by supernovae expanding into the interstellar medium are the origin of the galactic cosmic rays.

Can we observe neutrinos pointing back at the accelerators of the galactic cosmic rays? The conversion of the $10^{50}\,$erg of energy into particle acceleration is believed to occur by diffusive shock acceleration in the young (1000--10,000 years) remnant expanding into the interstellar medium. If 
high-energy cosmic rays are indeed associated with the remnant, they will interact with hydrogen atoms in the interstellar medium to produce pions that decay into photons and neutrinos. These may provide us with indirect evidence for cosmic ray acceleration. The observation of these pionic gamma rays has been one of the motivations for neutrino as well as ground-based TeV-energy astronomy. 
 
The HESS telescope opened a new era in astronomy by producing the first resolved images of sources in TeV gamma rays, particularly, in this context, of the supernova remnant RX J1713.7-3946\,\cite{hess}. The observed flux suggests that HESS may have identified the first site where protons are accelerated to energies typical of the main component of the galactic cosmic rays. Although the resolved image of the source reveals TeV gamma ray emission from the whole supernova remnant, it shows a clear increase of the flux in the directions of known molecular clouds. This is suggestive of protons, shock accelerated in the supernova remnant, interacting with the dense clouds to produce neutral pions that are the source of the observed increase of the TeV photon signal. The image shows filaments of high magnetic fields consistent with the requirements for acceleration to the energies observed. Follow-up observations of the source in radio-waves and X-rays failed to identify the population of electrons required to generate TeV photons by purely electromagnetic processes; for a detailed discussion see Ref.~\citenum{hiraga}.
 
Fitting the observed spectrum by purely electromagnetic processes is challenging because the relative height of the inverse Compton and synchrotron peaks requires very low values of the $B$-field, inconsistent with those required to accelerate the electron beam to energies that can accommodate the observation of 100\,TeV photons. Nevertheless, an exclusively electromagnetic explanation of the non-thermal spectrum is certainly not impossible and even favored by some \,\cite{Katz}. One can, for instance, partition the remnant in regions of high and low magnetic fields that are the respective sites of acceleration and inverse Compton scattering.

A similar extended source of TeV gamma rays tracing the density of molecular clouds has been identified near the galactic center. Protons apparently accelerated by the remnant HESS J1745-290 diffuse through nearby molecular clouds to produce a signal of TeV gamma rays that trace the density of the clouds\,\cite{GC}. Detecting this source in neutrinos will be challenging because it is relatively weak --- its TeV luminosity is only of order 0.1 Crab. This is presumably due to the larger distance to the source as compared to RX J1713.7-3946\,\cite{kistler,kappes}. Moreover, its large angular size on the sky results in a larger background of atmospheric neutrinos. So far, HESS has not claimed the discovery of pionic gamma rays and finding neutrinos as a smoking gun for cosmic ray acceleration in supernova remnants remains of interest.

Supernovae associated with molecular clouds are a common feature of associations of OB stars that exist throughout the galactic plane. Although not visible by HESS, intriguing evidence has been accumulating for the production of cosmic rays in the Cygnus region of the Galactic plane from a variety of experiments:
\begin{itemize}
\addtolength{\itemsep}{-.3cm}

\item The observation of the Cygnus region by the HEGRA air Cherenkov telescope resulted in the serendipitous discovery of a TeV $\gamma$-ray source\cite{Aharonian:2002ij} TeV J2032+4130, with an average flux of ${\sim}3\%$ of the Crab Nebula\cite{Crab} and a hard spectrum. Especially intriguing is its possible association with Cygnus OB2, a cluster of at least 2700 (identified) young, hot stars with a total mass of ${\sim} 10^4$ solar masses\cite{Knodlseder:2000vq}.  At a relatively small distance of approximately 1\,kpc, this is the largest massive Galactic stellar association.
\item The Whipple Observatory\cite{Lang:2004bk} confirmed an excess in the same direction as\break J2032+4130, although with considerably larger average flux above a peak energy response of 0.6~TeV.  Their latest results\cite{Konopelko:2006jr} reveal a TeV hot spot (integral flux ${\sim} 8\%$ of the Crab) that is displaced about 9 arc\,minutes to the northeast of the TeV J2032+4130 position.
\item A re-analysis\cite{Butt:2006js} of the radio surveys of the region revealed a weak non-thermal shell-like supernova remnant-type object with location and morphology very similar to the HEGRA source. It could be the cosmic ray engine that powers the OB association.
\item The Milagro Collaboration reports an excess of events from the Cygnus region at the $10.9\sigma$ level\cite{Abdo:2006fq}. The observed flux within a $3^\circ \times 3^\circ$ window centered at the HEGRA source is 70\% of the Crab at the median detected energy of 12~TeV. Such a flux largely exceeds the one reported by the HEGRA Collaboration, implying that there could be a population of unresolved TeV $\gamma$-ray sources within the Cygnus OB2 association. In fact, they report a hot spot, christened MGRO J2019+37, at right ascension $= 304.83^\circ \pm 0.14_{\rm stat} \pm 0.3_{\rm sys}$ and declination $= 36.83^\circ \pm 0.08_{\rm stat} \pm 0.25_{\rm sys}$\cite{Abdo:2006fq}. A fit to a circular 2-dimensional Gaussian yields a width of $0.32 \pm 0.12$ degrees, which for a distance of 1.7~kpc suggests a source radius of about 9~pc. The brightest hotspot in the Milagro map of the Cygnus region, it represents a flux of 0.5 Crab above 12.5~TeV. Interestingly, the Tibet AS-gamma Collaboration has observed a cosmic ray anisotropy from the direction of Cygnus, which is consistent with Milagro's measurements\cite{Amenomori:2006bx}.
\end{itemize}

As for the HESS sources, the observations suggest the production of cosmic rays as well as a variety of opportunities for neutrino production. The model proposed\cite{aongus} for MGRO J2019+37 is that of a cosmic ray beam which escapes from the OB star cluster and interacts with a molecular cloud positioned a few degrees to the southeast.

The discussion emphasizes the importance of observing the neutrinos from the decay of charged pions that accompany the TeV photons if these are indeed the decay products of neutral pions produced by cosmic rays interacting in the interstellar medium in the vicinity of the supernova remnant. The particle physics is simple: After oscillations over cosmic distances the neutrino beam consists of equal fluxes of muon, electron and tau-neutrinos and their antiparticles. Since proton-proton collisions yield two charged pions for every neutral pion, oscillations cause the flux for each neutrino flavor to equal one half of the gamma ray flux. Because the protons transfer on average 0.2 of their energy to secondary pions, and the four leptons in the charged pion decay chain $\pi \rightarrow \mu (\rightarrow e + \nu_{e} + \nu_{\mu}) + \nu_{\mu}$ take roughly equal energy, neutrinos with .05 of the cosmic ray energy are produced. Similarly,  photons with 0.1 of the proton energy are made from the decay of neutral pions. Accelerators producing cosmic rays reaching the ``knee" must produce photons with energies up to 100\,TeV and neutrinos up to half that energy. This requirement is consistent with observations of RX J1713.7-3946 and MGRO J2019+37 discussed above.

With 677 optical sensors in place since February 2000, the existing AMANDA detector has been collecting neutrinos at a steady rate of four per day. These atmospheric neutrinos are the byproduct of collisions of cosmic rays with atmospheric nitrogen and oxygen nuclei in the northern atmosphere. Note that at the South Pole one observes neutrinos that originate in the northern hemisphere, using the earth as a filter to select neutrinos from other particles. No photons, or any other particles besides neutrinos, can traverse the whole planet to reach the detector. The signals from the atmospheric neutrinos do not yield information about astronomy yet, but they are calculable and have been used to calibrate the detector. As in conventional astronomy, AMANDA looks through the atmosphere for cosmic signals and the data are scrutinized for hot spots in the northern sky that may signal cosmic sources. The fluxes from the sources discussed in this paper are small and, as has been known for some time, their detection requires the construction of neutrino detectors of kilometer scale.

Starting in the Antarctic summer 2004-2005, IceCube deployments have been steadily augmenting the AMANDA instrumentation. As of January 2007 IceCube consists of 1424 digital optical modules distributed over 22 strings and 54 surface cosmic ray detectors.  The hardware and software have worked Òout of the boxÓ and revealed the first atmospheric neutrinos. The instrumented volume of IceCube already exceeds one quarter of a kilometer cubed and will deliver a kilometer-square year of integrated data by 2008-09 provided the deployments remain on schedule. We will show that this represents the first opportunity to observe the muon neutrinos possibly accompanying the gamma rays observed in the Cygnus region. IceCube has sensitivity to the Southern HESS sources by observing the showers initiated by electron and tau neutrinos. A quantitative estimate is not possible as the fraction of these events for which the direction can be reconstructed with degree accuracy has not been published.

With the initial deployment and operation of strings in the Mediterranean, the Antares collaboration has observed its first neutrinos and is establishing a technology for constructing a kilometer-scale observatory in the Northern hemisphere. It will be ideally positioned to observe the supernova sources in the Southern part of the galactic plane including RX J1713.7-3946, the most promising source so far.

In this paper we will first review the qualitative arguments why the observed TeV gamma ray fluxes from the supernova remnants discussed above are consistent with those expected from a generic source building the steady galactic flux observed. We will argue that although the estimates are very uncertain, observation of the accompanying neutrino flux requires kilometer-scale detectors. Uncertainties in the calculation are associated with the propagation of the cosmic rays, with the value of the magnetic fields and the age of the remnant. Additionally, in the case of MGRO J2019+37, the spectral slope has not been measured. After investigating the wide parameter space of models for  MGRO J2019+37, we show that the neutrino flux can be predicted up to a factor of 2 once we match the fluxes at 12.5\,TeV to the Milagro data and limit the GeV flux by imposing the constraint that EGRET did not observe a GeV counterpart\cite{beacom}. This represents our main result. Using the published sensitivity of IceCube to muon neutrinos\cite{ice3} we predict $2\sim4$ events per year in a degree angular bin centered on MGRO J2019+37 on a background of atmospheric neutrinos of 2.5 events per year. Confirmation of cosmic ray acceleration should emerge after a few years of data taking.

\section{Secondary Pionic Gamma and Neutrino Fluxes from\hfil\break
\hbox to 1em{\hfil} Supernova Remnants}

A simple estimate is sufficient to conclude that the observed gamma ray flux associated with the supernova remnant RX J1713.7-3946 and MGRO J2019+37 is consistent with the energetics required from typical sources of galactic cosmic rays. The emissivity in pionic gamma rays (photons produced per cm$^{3}$ s) resulting from a density of accelerated protons $n_p$ interacting with a density of hydrogen atoms $n$ in the interstellar medium is
\begin{eqnarray}
Q_\gamma ({>} 1\,{\rm TeV})& =& c \left<{E_\pi \over E_p}\right> \sigma_{pp}\, n\, n_p\ ({>}1\,{\rm TeV})\\
&=& c \left<{E_\pi \over E_p}\right> {\lambda_{pp}}^{-1}\, n_p\ ({>}1\,{\rm TeV})\\
\noalign{\hbox{or}}
Q_\gamma (> 1\,{\rm TeV}) &\simeq& 10^{-29} \,{\rm photons\over \rm cm^3\,s}\, \left({n \over \rm 1\,cm^{-3}}\right).
\end{eqnarray}
The emissivity of photons is simply proportional to the density of cosmic rays\break $n_p(>1\,{\rm TeV})$ ($\simeq4\times10^{-14}\, {\rm cm}^{-3}$ for energy in excess of 1 TeV) and the target density $n$ of hydrogen atoms. The proportionality factor is determined by particle physics: $\left<E_\pi/E_p\right> \sim 0.2$ is the average energy of the secondary pions relative to the cosmic ray protons and $\lambda_{pp}= (n\sigma_{pp})^{-1}$ is the proton interaction length ($\sigma_{pp} \simeq 40$\,mb) in a density $n$ of hydrogen atoms. (We here assumed a generic $E^{-2}$ spectrum of the protons, for different spectral indices the quantity $\left<E_\pi/E_p\right>$ is generalized to the spectrum-weighted moments for pion production by nucleons\cite{gaisser}; see later).

The total luminosity in gamma rays is given by
\begin{equation}
L_{\gamma} ({>} 1\,{\rm TeV})  = Q_\gamma\, {W \over \rho} \simeq 10^{33}\, \rm photons\> s^{-1}.
\end{equation}
The density of protons from a supernova converting a total kinetic energy $W$ of $10^{50}$\,erg to proton acceleration is approximately given by $W/\rho$, where we will assume that the density in the remnant is not very different from the ambient energy density $\rho \sim 10^{-12}$\,erg\,cm$^{-3}$ of galactic cosmic rays. This approximation is valid for young remnants in their Sedov phase\cite{ADV94}.

We thus predict a  rate of TeV photons from a supernova at a distance $d$ of 1\,kpc~of
\begin{equation}
{dN_{\rm events}\over d({\mit ln}E)}\,({ >}E) = {L_\gamma \over 4\pi d^2} \simeq 10^{-11}  \left({\rm photons\over \rm cm^2\,s}\right) \left({W\over \rm 10^{50}\,erg}\right) \left({n\over \rm 1\,cm^{-3}}\right) \left({d\over \rm 1\,kpc}\right)^{-2}.
%\label{galactic1}
\end{equation}
This prediction is credible because the number of TeV photons predicted coincides with observations of the supernova remnant RX J1713.7-3946 by the HESS array of atmospheric Cherenkov telescopes\cite{hess}.

Within the precision of the astrophysics it is adequate to assume that muon neutrinos and muon antineutrinos are produced at the level of one half the gamma flux as explained in the introduction. We thus anticipate an event rate of ${\sim}1.5$ detected neutrinos per decade of energy per km$^2$ year, a result readily obtained from the relation
\begin{equation}
{dN_{\rm events}\over d(ln\,{\rm E})}\, ({>}E) = 10^{-11} \: \left({\rm neutrinos\over \rm cm^2 \: s}\right) \: \rm area \: \rm time \: \left({\lambda_\mu\over \rm \lambda_\nu}\right),
\label{galactic2}
\end{equation}
where the last factor represents, as before, the probability that the neutrino is detected. It is approximately $10^{-6}$ for the TeV energy considered here. This estimate may be somewhat optimistic because we assumed that the sources extend to 100\,TeV with an $E^{-2}$ spectrum. The number of neutrinos is uncomfortably small and suggests a more sophisticated calculation that allows for an estimate of the range of fluxes allowed within observational constraints. We perform such an analysis for the observation of the Cygnus region by IceCube, whose acceptance is simulated following 
Refs.~\citenum{Gonzalez-Garcia:2005xw} and \citenum{Ahrens:2003ix}.

Alternative candidates have been suggested for the sources of the galactic cosmic rays for, instance microquasars. The above argument suggests that they should have left their imprint on the Milagro skymap and they did not. It is very suggestive that the Milagro sources are the cosmic ray accelerators.

\section{Remnants in the Sedov Phase: Fluid Effects on the Proton Spectrum}

The excess of events detected by the Milagro Collaboration in the Cygnus region of the Milky way
may be due to a shell-type supernova remnant (SNR). In such an object, a spherical shock sweeps
through the local ISM, colliding charged particles accelerated by Fermi shock acceleration into the
ambient medium. These collisions lead to the production of neutral and charged pions that decay 
to photons and leptons, respectively, as discussed in Section 2. 

The injected proton spectrum $Q(E)$ will be modified by fluid processes acting on the particles as they 
traverse the volume of the SNR during acceleration. Normally, considering only the shock 
acceleration gives some probability for particle escape during each successive traversal of the
shock front. We consider two additional mechanisms that allow the protons to escape and 
thereby modify the final proton spectrum: convection and diffusion. Convection is the 
escape of particles due to currents within the object, while diffusion represents the random walk of particles resulting in a net motion down the particle density gradient. Since the mean free path scales with energy, diffusion will steepen the high-energy tail of the proton spectrum, with high energy particles escaping before those with lower energies. Here we assume Bohm diffusion, which gives us the dependence on energy of the escape time. 

We take the equation describing the final proton spectrum $N_{p}$ to be; see e.g.\ Ref.~\citenum{crocker}
\begin{equation}
\frac{dN(E)}{dt}=Q(E) - \frac{N(E)}{\tau(E)},
\end{equation}
where $\tau(E)$ is the characteristic escape time of the protons. We find 
the solution in the steady-state limit as
\begin{equation}
N(E) = Q(E)\,\tau(E)\,,
\end{equation}
where $\tau(E)=\tau_{\rm fluid}(E)$ when $\tau_{\rm fluid}(E)<t_{\rm SNR}$ and $\tau(E)=t_{\rm SNR}$
when $\tau_{\rm fluid}(E)>t_{\rm SNR}$; $t_{\rm SNR}$ is the age of the supernova remnant and 
$\tau_{\rm fluid}(E)$ is the escape time due to fluid effects.  

$\tau_{\rm fluid}(E)$ is given by the harmonic mean of the convective and diffusive escape times
\begin{equation}
\frac{1}{\tau_{\rm fluid}} = \frac{1}{\tau_{\rm diff}} + \frac{1}{\tau_{\rm conv}}, 
\end{equation}
where 
\begin{equation}
\tau_{\rm conv} = \frac{kR}{v_{\rm shock}}
\end{equation}
and
\begin{equation}
\tau_{\rm diff} = (4524~{\rm years}) \times \left(\frac{B}{\mu \rm Gauss}\right) \times \left(\frac{R}{pc}\right)^{2}\times \left(\frac{E_{p}}{1~\rm TeV} \right)^{-1}
\end{equation}
for $k$ the compression factor of the gas, $R$ and $v$ the radius and velocity of the 
shock, and $B$ the magnetic field in the region. The parametrization of $\tau_{\rm diff}$ is from 
Ref.~\citenum{lemoine} and assumes Bohm diffusion, hence the $E^{-1}$ scaling. We choose and hold constant $R=10\,{\rm pc}$ and ${\rm v_{\rm shock}}=1000\,{\rm km/s}$. While above approximations may be inadequate for a detailed study of the sources, they are sufficient for deriving the neutrino flux from the astronomical observations. The calculation is insensitive to the detailed astronomical parameters as we will show further on.

\begin{figure}[h]
\begin{center}
\epsfig{file=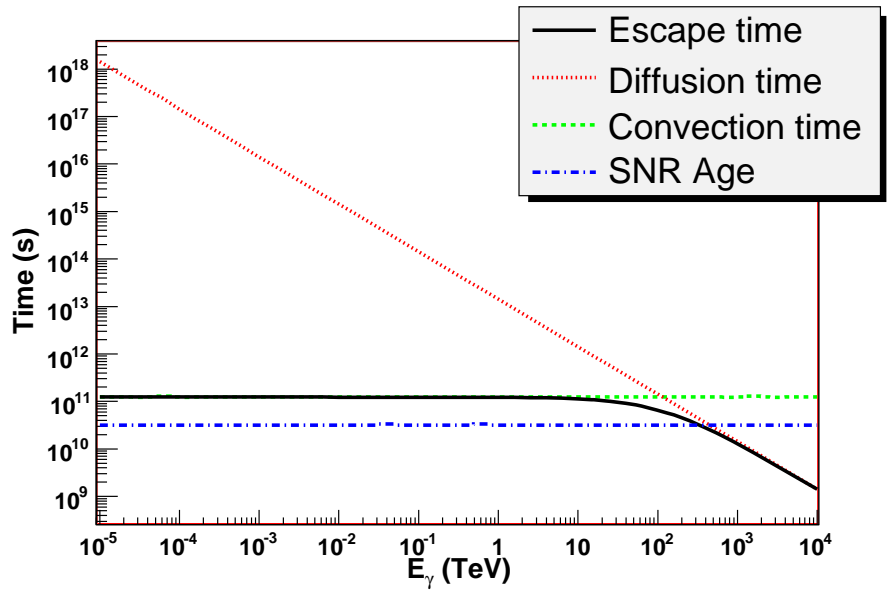, scale=0.6}
\end{center}
\caption[times]{Comparison of SNR age (blue dash-dotted), fluid escape time (solid black), diffusion time (red dotted) 
and convection time (green dashed) for a sample parameter set of ${\rm age}\,=\,1000$ years, $B-$field${}=\,1\,\mu$G, ${\rm v_{\rm shock}}\,=\,1000\,{\rm km/s}$ and $R\,=\,10\,{\rm pc}$ . In our model we modify the spectrum in proportion to whichever of the age or the fluid escape time is lower at a particular energy.}
\label{fig:times}
\end{figure}

The result of modifying the initial proton spectrum by these mechanisms is equivalent to 
introducing a gradual cutoff at the energy where the $E^{-1}$ behaviour of the Bohm diffusion 
takes hold--essentially, $E^{-1}$ is superimposed on the injected spectrum. In the limits we describe above, this transition energy is determined by the intersection of the fluid escape time curve with the age curve of the SNR (Figure~\ref{fig:times}). We modify the spectrum by multiplying it by whichever of the two times represents a smaller escape time at each energy. This means that if $\tau_{\rm fluid}>t_{\rm SNR}$ for a given energy, the spectrum is simply multiplied by some constant to be determined by the normalization and its shape at that energy is unmodified from the original injection spectrum. While convective transport is included in the model for completeness, the characteristic time for convective escape does not depend on energy. Therefore, varying the convection parameters as well as the age of the remnant is redundant as their effects on the final gamma-ray spectrum are almost identical up to normalization. Hence the holding constant of the convection parameters.

%Fig.1

This model introduces a number of parameters whose precise values for the Cygnus object are unknown although there is a region of the parameter space considered most physically likely: radii of order
several parsecs, velocities of order several tens of thousands of km/s, magnetic fields of order between
several and several tens of microGauss, and ages between 500 and 10,000 years. However, since we are normalizing the resulting
$\gamma$-ray spectrum to the observations made in ${\sim}$10\,TeV \mbox{$\gamma$-rays} by the Milagro experiment and 
constraining 
the spectrum to be below the fluxes measured by the EGRET satellite experiment in the GeV range,
the model becomes highly degenerate. There are essentially two regimes: 
\begin{enumerate}
\addtolength{\itemsep}{-.2cm}

\item Diffusion begins to 
affect the spectrum at energies higher than the Milagro observation at 12\,TeV. This happens when the
$B$-field is large or the age of the object is small. In this case, the loss of particles will happen
at an energy at which the flux was already very low, resulting in little effect on the neutrino
event rates from an undiffused spectrum; 
\item Diffusion begins to affect the spectrum at energies lower
than the Milagro point. This means that the Milagro observation is of a flux already reduced by this fluid 
mechanism, resulting in a potentially large boost to the flux in the GeV range. However, the largest boosts to the midrange energies 
can only be obtained by considering potentially unphysical values of the fluid parameters --- $B$ of order several tenths or hundreths of a microGauss, for example. This range is also constrained by the EGRET measurements.
\end{enumerate}

\section{Determining the Gamma-Ray Spectrum from the Proton Spectrum}

In this section we discuss a model of the proton acceleration that allows us to derive a pionic
gamma-ray flux. We assume a shock accelerated proton input population that can be described by
a power law with an exponential cutoff. To accommodate the ``knee'' in the cosmic-ray spectrum, 
we put the cutoff at 1\,PeV:
\begin{equation}
Q_{p} = A\,E^{s}\,e^{-E/1\,{\rm PeV}}\,.
\end{equation}
The choice of cutoff, although chosen to be at the same scale as the knee at 3\,PeV, is not critical to the following analysis. The flux is so low at PeV energies that there are zero events regardless of the precise location of the cutoff.  

The basic formula for the gamma spectrum is
\begin{equation}
\frac{dN_{\gamma}}{dE} = c\,n\,\int_{E_{\gamma}}^{\infty}\, N_{p}(E_{p})\, F(E_{p}, E_{\gamma})\, \sigma(E_{p})\, \frac{dE_{p}}{E_{p}}\,,
\end{equation}
where $N_{p}$ is the proton spectrum modified by fluid effects (previous section) and $F$ is a 
fragmentation function defined in Ref.~\citenum{KAB}
describing the energy transfer of the accelerated protons to neutral mesons (pions and etas) and 
subsequently, to $\gamma$-rays. 

We plotted the gamma-ray flux predicted by the above model for a range of parameters, normalized
to the Milagro observation at 12.5\,TeV and constrained by the EGRET observations in the 
$\sim$100\,MeV--10\,GeV energy range. Following Ref.~\citenum{kistler} the EGRET source 3EG J2021+3716 provides the lower-energy constraint. It is clear that 3EG J2021+3716 is either the counterpart to MGRO J2019+37, or, if it is not, the true counterpart must have a smaller flux than it in order to have escaped detection. In either case, the EGRET measurement gives us a clear upper limit in the low-energy $\gamma$-ray range.  

\begin{figure}[h]
\begin{center}
\epsfig{file=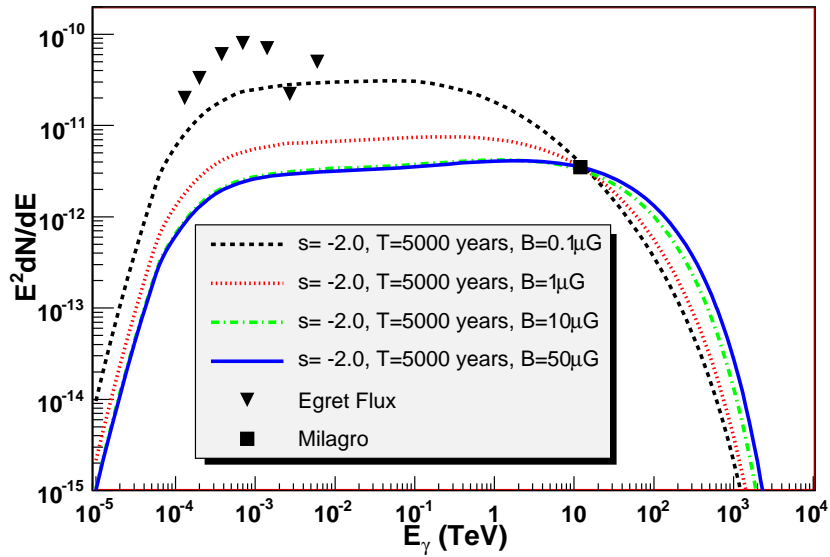, scale=0.6}
\end{center}
\caption[20spectrum]{$\gamma$-ray spectra with injection $s=-2.0$. All lines are for 5000 year old SNR. The black dashed line is for a magnetic field of 0.1~$\mu$G, the red dotted line is for a magnetic field of 1~$\mu$G, the green dash-dotted line is for a magnetic field of 10~$\mu$G, and the blue solid line is for a magnetic field of 50~$\mu$G. The inverted triangles are show the EGRET measurement of 3EG J2021+3716 and the solid box is the Milagro measurement of MGRO J2019+37.}
\label{fig:20spectrum}
\end{figure}

\begin{figure}[h!]
\begin{center}
\epsfig{file=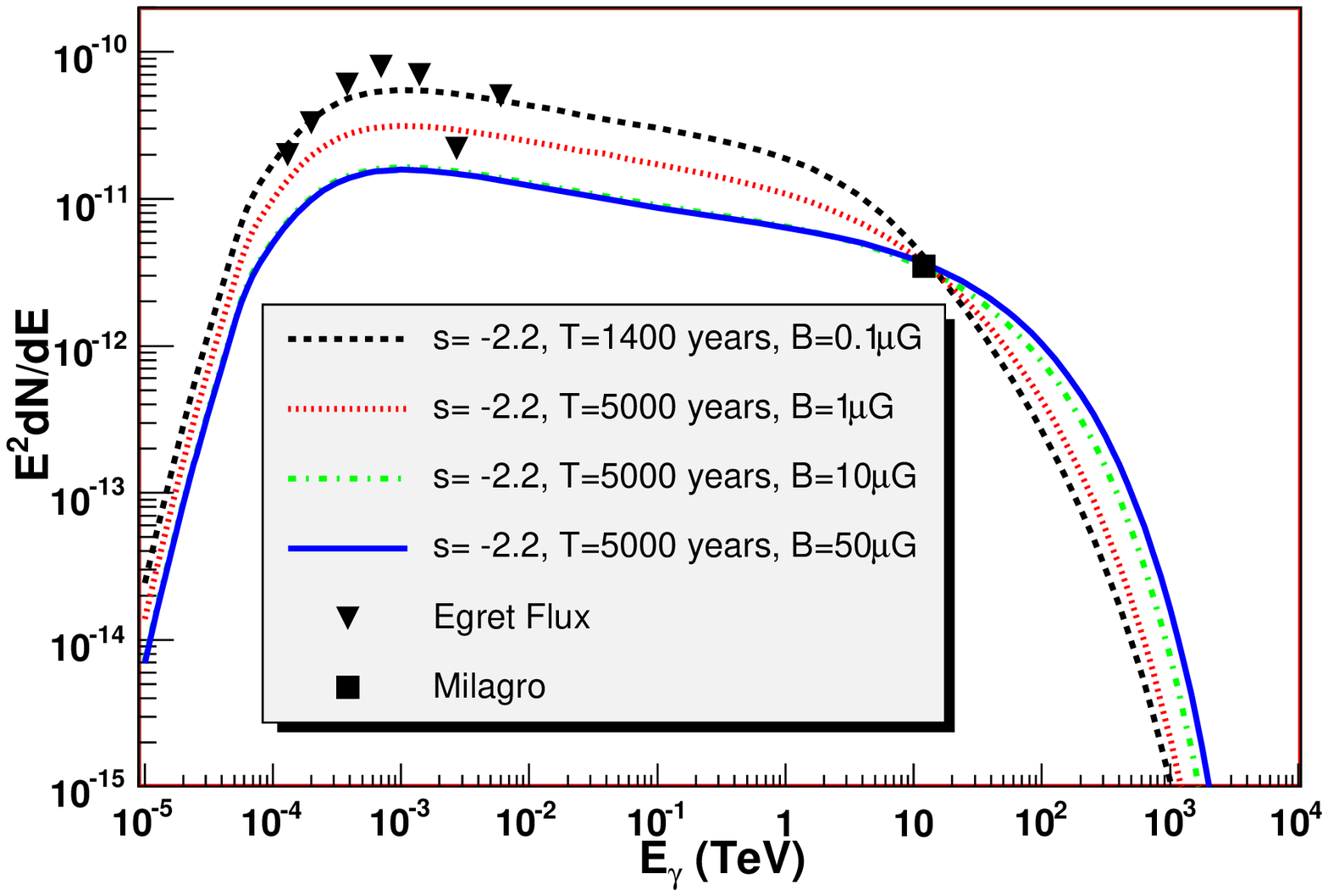, scale=0.6}
\end{center}

\vspace*{-.5cm}

\caption[22spectrum]{$\gamma$-ray spectra with injection $s=-2.2$. The black dashed line is for a magnetic field of 0.1~$\mu$G and age of 1400\,years, the red dotted line is for a magnetic field of 1~$\mu$G and age of 5000\,years, the green dash-dotted line is for a magnetic field of 10~$\mu$G and age of 5000\,years, and the blue solid line is for a magnetic field of 50~$\mu$G and age of 5000\,years.}
\label{fig:22spectrum}
%\end{figure}

\vspace{1cm}

%Fig.4
%\begin{figure}[b]
\begin{center}
\epsfig{file=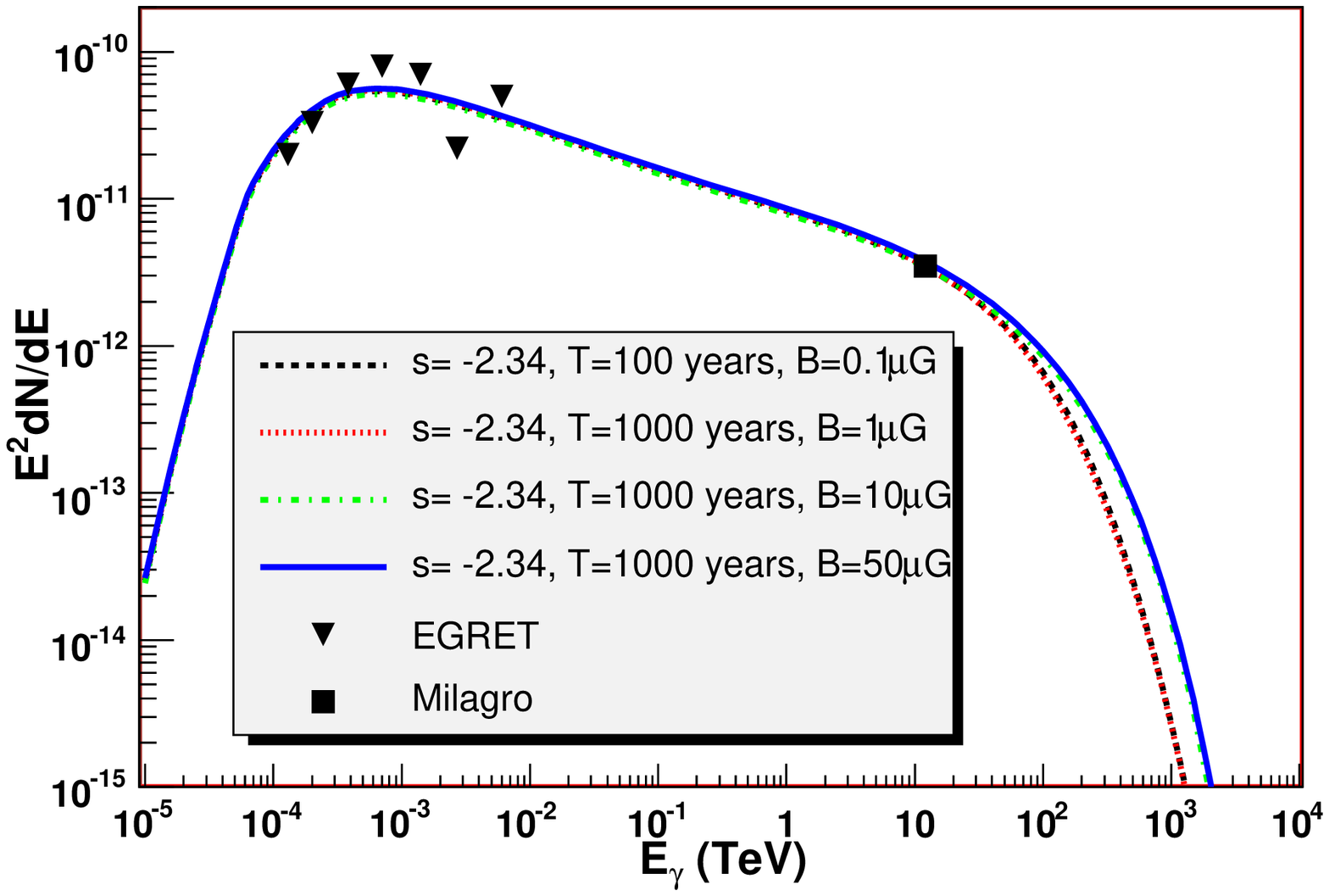, scale=0.6}
\end{center}

\vspace*{-.5cm}

\caption[234spectrum]{$\gamma$-ray spectra with injection $s=-2.34$. The black dashed line is for a magnetic field of 0.1~$\mu$G and age of 100\,years, the red dotted line is for a magnetic field of 1~$\mu$G and age of 1000\,years, the green dash-dotted line is for a magnetic field of 10~$\mu$G and age of 1000\,years, and the blue solid line is for a magnetic field of 50~$\mu$G and age of 1000\,years.}
\label{fig:234spectrum}
\end{figure}

We took three sample input proton spectra $s=-2.0, -2.2, -2.34$ and plotted
each for four values of the magnetic field ranging from 0.1 to 50 $\mu G$. The age of the SNR was
then chosen to be the value that maximized the gamma-ray flux for the previously chosen set of 
parameters, subject to the EGRET constraint (Figures~\ref{fig:20spectrum}--\ref{fig:234spectrum}). For $s > -2.34$, there were no gamma-ray spectra that did not exceed the EGRET upper limit.  Indeed, $s=-2.34$ (Figure~\ref{fig:234spectrum}) proved such a constraint on the fluid parameter space that the variation in the allowed spectra produces only a very small variation in neutrino event rates in IceCube.  Due to the degeneracy of the model discussed earlier, we did not need to vary the variables associated with convection. 

%% Fig.2

%\floatfraction{1}

%Fig.3

\newpage

\section{Neutrino Event Rates}
We calculated the neutrino flux $dN_{\nu}/dE_{\nu}$ from the gamma flux by the method of 
Ref.~\citenum{cavasinni}, 
assuming the two are equal up to some constant that also includes the effects of oscillations (see 
Ref.~\citenum{bolofoot} for an alternative method and results). We can neglect 
the energy dependence of the oscillations due to the large distance to the source. Given a neutrino flux, the event rate in IceCube is simply the convolution 
of the flux with the energy-dependent area and the probability of an incident neutrino inducing a muon with a visible track. The area $A_{\rm eff}$ is 
taken to be the effective area after quality cuts on the IceCube data referred to as level 2 cuts in
Ref.~\citenum{Ahrens:2003ix}, 

\begin{equation}
N_{\nu} = T\int_{E_{\rm thresh}}\,A_{\rm eff}(E_{\nu})\, \frac{dN_{\nu}}{dE_{\nu}}\,P_{\nu\rightarrow\mu}\,dE_{\nu}\,.
\end{equation}

\begin{figure}[htbp]
\begin{center}
\epsfig{file=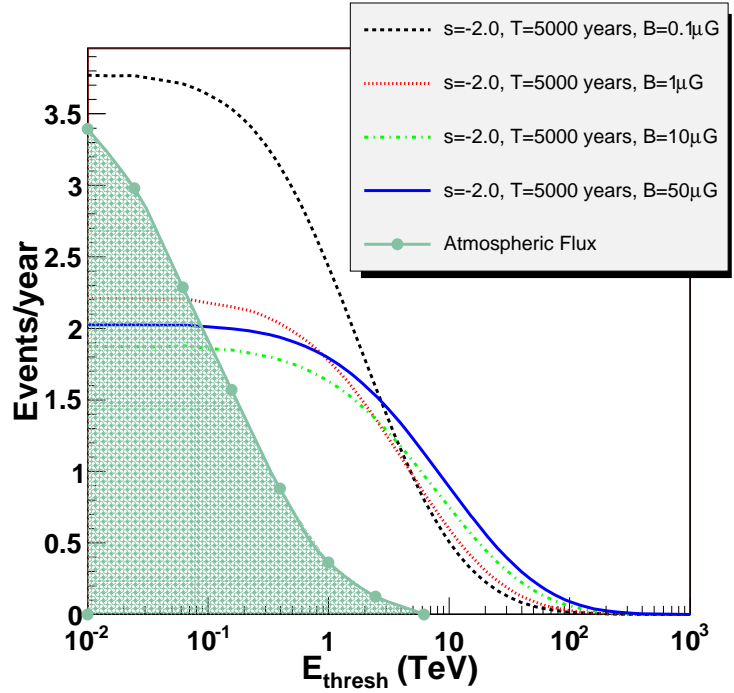, scale=0.5}
\end{center}
\caption[20events]{Events due to the gamma-ray spectra with injection $\alpha=-2.0$ shown in Fig.~\ref{fig:20spectrum}. }
\label{fig:20events}
\end{figure}

\begin{figure}[htbp]
\begin{center}
\epsfig{file=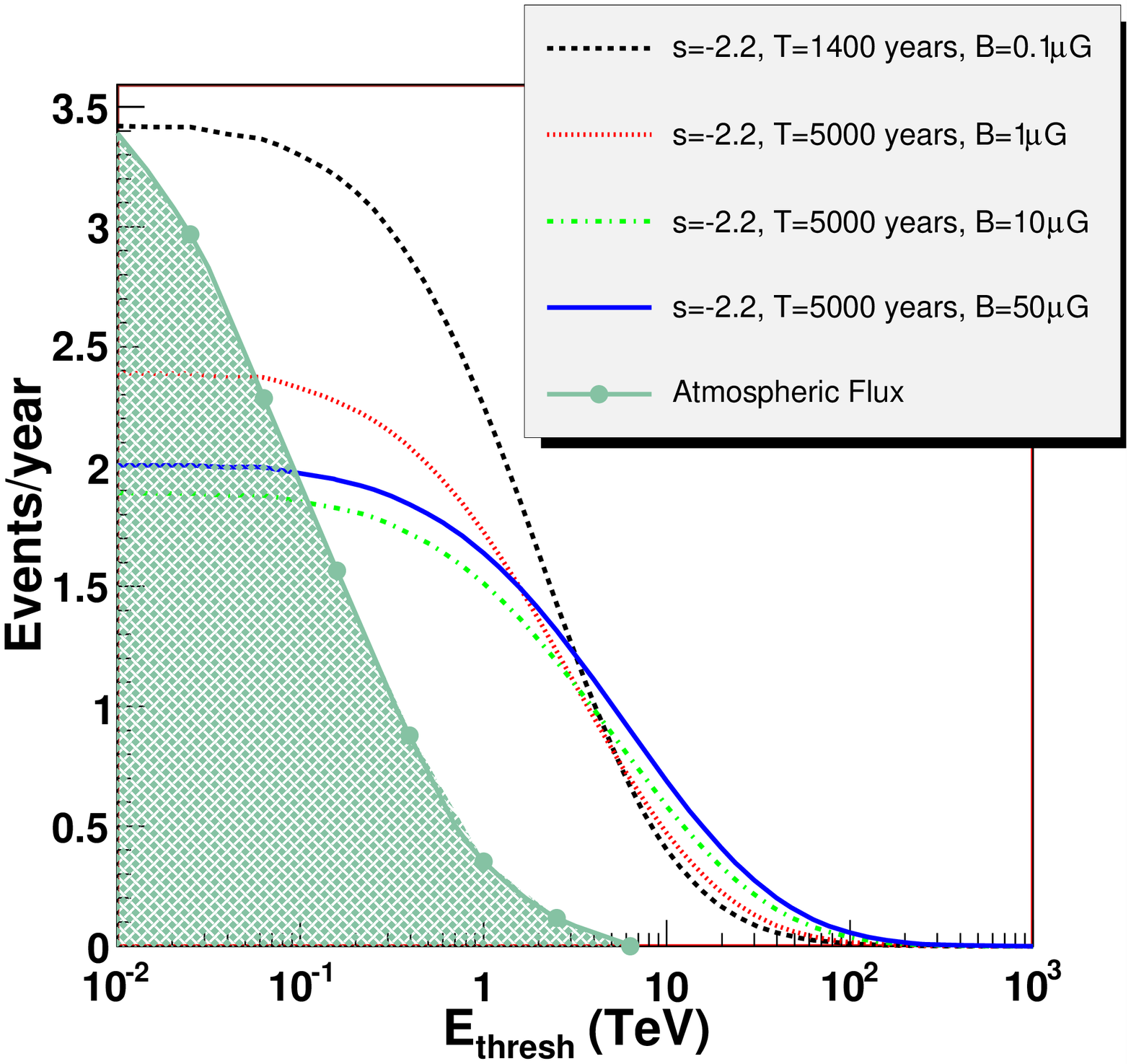, scale=0.5}
\end{center}
\caption[22events]{Events due to the gamma-ray spectra with injection $\alpha=-2.2$ shown in Fig.~\ref{fig:22spectrum}.}
\label{fig:22events}
\end{figure}

\begin{figure}[htbp]
\begin{center}
\epsfig{file=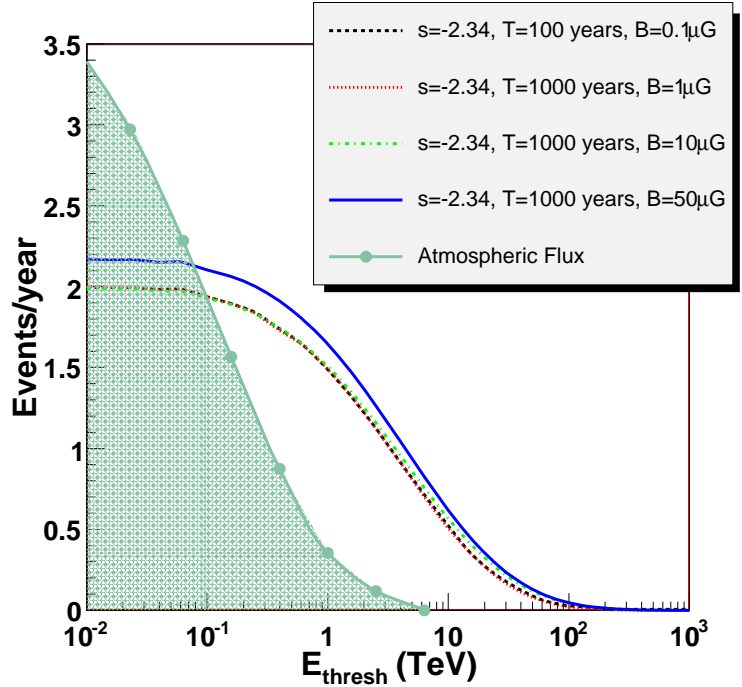, scale=0.5}
\end{center}

\vspace*{-.5cm}

\caption[234events]{Events due to the gamma-ray spectra with injection $\alpha=-2.34$ shown in Fig.~\ref{fig:234spectrum}.}
\label{fig:234events}

\end{figure}

The event rates in IceCube are calculated in this way are shown in Figures~\ref{fig:20events}--\ref{fig:234events}.
The rates are within the range $2\leq dN/dt \leq 3.8$ events per year with the IceCube threshold at 50\,GeV.
This is in large part due to the fact that the Milagro observation strongly constrains the flux in the
energy range 1--20\,TeV, where the neutrino detection probability is highest, resulting in similar predictions for dissimilar SNR characteristics. The irreducible 
atmospheric background, due to neutrinos produced in the Northern atmosphere in cosmic ray showers, is
calculated using the results of Ref.~\citenum{agrawal} and is found to be approximately 2.5 neutrinos/year at 50\,GeV. Hence in 15 years of operation we predict 
$4.9~\sigma \leq~N/\sqrt{N_{\rm atmo}}~\leq 9.3~\sigma$ and if the higher end of the predicted event rate range is realized, $5~\sigma$ is possible in 4.3 years. 

In closing, we note that the Milagro collaboration\cite{sinnis} has recently detected multiple additional sources besides MGRO J2019+37, most with approximately equal fluxes of 0.5 Crab. The sources with possible counterparts in the GeV range indicate a spectral index of ${\sim}-2.3$, bearing out our observation in Section~4 that spectral indices steeper than $-2.34$ exceed the EGRET signal from the possible GeV counterpart to MGRO J2019+37 for all model parameters.  If we compute the flux of neutrinos from the Milagro sources (not including the Crab Nebula) detected with post-trial significance of greater than five $\sigma$ assuming a power-law index of $-2.34$, we get a total event rate in IceCube of 6.9 neutrinos/year.  If we also include the more tentative sources, the event rate increases to 11.5 neutrinos/year. In the long run, a correlation analysis of the IceCube and Milagro skymaps should make the detection of these sources likely. 

\newpage

\section{Acknowledgments}
We thank our IceCube collaborators as well as Luis Anchordoqui, Julia Becker, Tom Gaisser, Concha Gonzalez-Garcia, Jordan Goodman, Teresa Montaruli, Greg Sullivan and Grant Teply for discussions. This research was supported in part by the National Science Foundation under Grant No.~OPP-0236449, in part by the U.S.~Department of Energy under Grant No.~DE-FG02-95ER40896, and in part by the University of Wisconsin Research Committee with funds granted by the Wisconsin Alumni Research Foundation.

%\vspace*{-.75cm}


\begin{thebibliography}{99}
%
%\vspace*{-.5cm}

\bibitem{hess}
F.A. Aharonian {\it et al.}, Nature {\bf 432}, 75 (2004), arXiv:astro-ph/00411533; H. J. Volk, E.G. Berezhko, Leonid T. Ksenofontov, arXiv:astro-ph/0409453.

\bibitem{hiraga}
Y. Uchiyama {\it et al.} , International Symposium on High-Energy Gamma-ray Astronomy, Heidelberg, Germany (2004), arXiv:astro-ph/00503199.

\bibitem{Katz}
For a recent discussion, see B.~Katz and E.~Waxman, arXiv:astro-ph/0706.3485. 

\bibitem{GC}
F.A. Aharonian {\it et al.}, Nature {\bf 439}, 695 (2006), arXiv:astro-ph/0603021.


\bibitem{kistler}
J.~F.~Beacom and M.~D.~Kistler, arXiv:astro-ph/0607082.

\bibitem{kappes}
A. Kappes {\it et al.}, arXiv:astro-ph/0607286.

\bibitem{Aharonian:2002ij}
  F.~A.~Aharonian, A.~Akhperjanian, M.~Beilicke, Y.~Uchiyama and T.~Takahashi,
  %``An unidentified TeV source in the vicinity of Cygnus OB2,''
  Astron.\ Astrophys.\  {\bf 393}, L37 (2002), arXiv:astro-ph/0207528.
  %%CITATION = ASTRO-PH 0207528;%%

\bibitem{Crab} The integral $\gamma$-ray flux obtained from the Crab
  by the Whipple Collaboration is now the standard TeV~flux unit:
  $F_{\rm Crab}(E_\gamma>0.35\,{\rm TeV})= 10^{-10}~{\rm cm}^2\, {\rm
    s}^{-1}$.  The spectral index of the Crab's integrated flux is
measured to be $-1.5$. G.~Vacanti {\em et al.},
Astrophys. J. 377, 467 (1991); A.~M.~Hillas {\em et al.}, Astrophys.
J. 503, 744 (1998).

\bibitem{Knodlseder:2000vq}
  J.~Knodlseder,
  %``Cygnus OB2 - a young globular cluster in the Milky Way,''
  arXiv:astro-ph/0007442.
  %%CITATION = ASTRO-PH 0007442;%%

\bibitem{Lang:2004bk}
  M.~J.~Lang {\it et al.},
  %``Evidence for TeV gamma ray emission from TeV J2032+4130 in Whipple
  %archival data,''
  Astron.\ Astrophys.\  {\bf 423}, 415 (2004), arXiv:astro-ph/0405513.
  %%CITATION = ASTRO-PH 0405513;%%

\bibitem{Konopelko:2006jr}
  A.~Konopelko {\it et al.},
  %``Observations of the unidentified TeV gamma-ray source TeV J2032+4130 with
  %the Whipple Observatory 10-m telescope,''
  arXiv:astro-ph/0611730.
  %%CITATION = ASTRO-PH 0611730;%%


\bibitem{Butt:2006js}
  Y.~M.~Butt, J.~A.~Combi, J.~Drake, J.~P.~Finley, A.~Konopelko, M.~Lister 
  and J.~Rodriguez,
  %``TeV J2032+4130: A not-so-dark accelerator?,''
  arXiv:astro-ph/0611731.
  %%CITATION = ASTRO-PH 0611731;%%

\bibitem{Abdo:2006fq}
  A.~A.~Abdo {\it et al.},
  %``Discovery of TeV gamma-ray emission from the Cygnus region of 
  %the galaxy,''
  arXiv:astro-ph/0611691.
  %%CITATION = ASTRO-PH 0611691;%%

\bibitem{Amenomori:2006bx}
  M.~Amenomori  [Tibet AS-gamma Collaboration],
  %``Anisotropy and corotation of galactic cosmic rays,''
  Science {\bf 314}, 439 (2006), arXiv:astro-ph/0610671.
  %%CITATION = ASTRO-PH 0610671;%%

\bibitem{aongus}
L. Anchordoqui, F. Halzen, T. Montaruli and A. O'Murchadha, arXiv:astro-ph/0612699.

\bibitem{beacom}
T.~Prodanovic, B.~D.~Fields and J.~F.~Beacom,
Astropart.\ Phys.\  {\bf 27}, 10 (2007), arXiv:astro-ph/0603618. 

\bibitem{ice3}
A.~Achtenberg  {\it et al.},  First Year Performance of the IceCube Neutrino Telescope, arXiv:astro-ph/0604450; J.~Ahrens  et al. (IceCube Collaboration), Astropart.\ Phys.\  {\bf 20}, 507 (2004), arXiv:astro-ph/0305196 and http://icecube.wisc.edu.
%

\bibitem{gaisser}
T. K. Gaisser, Cosmic Rays and Particle Physics, Cambridge University Press, Cambridge (2007).

\bibitem{ADV94}
F.~Aharonian, L.~O'C.~Drury and H.~J.~V\"{o}lk, {\it A\&A}, {\bf 285}, 645 (1994)
 



\bibitem{Gonzalez-Garcia:2005xw}
  M.~C.~Gonzalez-Garcia, F.~Halzen and M.~Maltoni,
  %``Physics reach of high-energy and high-statistics Icecube
  %atmospheric neutrino data,''
  Phys.\ Rev.\ D {\bf 71}, 093010 (2005), arXiv:hep-ph/0502223.
  %%CITATION = HEP-PH 0502223;%%

\bibitem{Ahrens:2003ix}
J.~Ahrens {\it et al.}  [IceCube Collaboration],
%``Sensitivity of the IceCube detector to astrophysical sources of high
%energy muon neutrinos,''
Astropart.\ Phys.\  {\bf 20}, 507 (2004), arXiv:astro-ph/0305196.
%%CITATION = ASTRO-PH 0305196;%%


\bibitem{crocker}
M.~Fatuzzo, F.~Melia, R.~M.~Crocker, arXiv:astro-ph/0602330.


\bibitem{lemoine}
M.~Lemoine-Goumard,
Ph.\ D thesis, L'Ecole Polythechnique, 2006.

\bibitem{KAB}
S.~R.~Kelner, F.~A.~Aharonian, V.~V.~Bugayov,
Phys.\ Rev.\ D{\bf 74}, 034018 (2006), arXiv:astro-ph/0606058.



\bibitem{cavasinni}
V.~Cavasinni, D.~Grasso, L.~Maccione,
Astropart.\ Phys.\ {\bf 26}, 41 (2006), arXiv:astro-ph/0604004.

\bibitem{bolofoot}
The semianalytical bolometric method outlined in Ref.~\citenum{alvarezhalzen}, where we assume equal energy in neutrinos as in gamma rays and equate 
the power fluxes, allows us to determine the neutrino flux using:
\begin{displaymath}
\int E_{\gamma} \frac{dN_{\gamma}}{dE_{\gamma}}dE_{\gamma} =\int E_{\nu} \frac{dN_{\nu}}{dE_{\nu}}dE_{\nu}
\end{displaymath}
While this method will determine the normalization of the neutrino spectrum, we must posit \textit{a priori}
a slope $s_{\nu}$ (assuming we take the neutrino spectrum as a simple power law) as well as 
an upper threshold. In this case we take $s_{\nu}=-2.0$ and $E_{\nu}^{\rm max} = 50$ TeV, recognizing
that in principle we cannot assume a pionic origin for the gamma rays or that the sequence of 
events that we have chosen to have evolve the spectra is the correct one. The main consequence of using this method is that the
variation in energy in the $\gamma$-ray flux at the EGRET energy range causes a large variation in the harder neutrino flux at 
GeV--TeV energies, potentially greatly increasing the number of detected neutrinos per year. By this method, we find event rates between
2 and 12 neutrinos/year, giving, respectively, 4.9 and 29.4 $\sigma$ after 15 years of operation. 

\bibitem{agrawal}
V.~Agrawal, T.~Gaisser, P.~Lipari, T.~Stanev,
Phys.\ Rev.\ D{\bf 53}, 1314 (1996), arXiv:hep-ph/9509423.

\bibitem{sinnis}
G. Sinnis for the Milagro Collaboration, VERITAS First Light Fiesta, Mt. Hopkins, Arizona (April 2007).



\bibitem{alvarezhalzen}
J. Alvarez-Muniz and F. Halzen, Ap.~J. {\bf 576}, L33 (2002).

\end{thebibliography}
\end{document}